\documentclass[amsmath,amssymb,twocolumn,prd,floatfix,showpacs, nofootinbib]{revtex4-1}
\usepackage{tabularx}  
\usepackage{bm, color} 
\usepackage{overpic,subfigure} 
\usepackage{multirow}
\usepackage{array}
\usepackage{dcolumn} 
\usepackage[symbol]{footmisc}
\usepackage{epstopdf}
\usepackage{ulem}
\RequirePackage{xspace}

\newcommand{\gev}{\ensuremath{\mathrm{\,Ge\kern -0.1em V}}\xspace}
\newcommand{\mev}{\ensuremath{\mathrm{\,Me\kern -0.1em V}}\xspace}
\newcommand{\mevcc}{\ensuremath{{\mathrm{\,Me\kern -0.1em V\!/}c^2}}\xspace}

\newcommand{\BESIIIorcid}[1]{\href{https://orcid.org/#1}{\hspace*{0.1em}\raisebox{-0.45ex}{\includegraphics[width=1em]{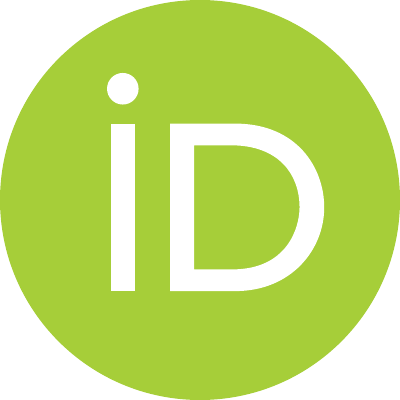}}}}

\def\fz#1       {\ensuremath{f_0({#1})}\xspace}

\usepackage{hyperref}
\hypersetup{colorlinks = true,
            linkcolor = black,
            urlcolor = blue,
            citecolor = blue}
\begin{document}
\title{\boldmath Observation and branching fraction measurements of $J/\psi \to p \bar p K^0_S K^0_S$ and $\psi(3686) \to p \bar p K^0_S K^0_S$}

\author{
	M.~Ablikim$^{1}$\BESIIIorcid{0000-0002-3935-619X},
	M.~N.~Achasov$^{4,c}$\BESIIIorcid{0000-0002-9400-8622},
	P.~Adlarson$^{82}$\BESIIIorcid{0000-0001-6280-3851},
	X.~C.~Ai$^{88}$\BESIIIorcid{0000-0003-3856-2415},
	C.~S.~Akondi$^{31A,31B}$\BESIIIorcid{0000-0001-6303-5217},
	R.~Aliberti$^{39}$\BESIIIorcid{0000-0003-3500-4012},
	A.~Amoroso$^{81A,81C}$\BESIIIorcid{0000-0002-3095-8610},
	Q.~An$^{78,65,\dagger}$,
	Y.~H.~An$^{88}$\BESIIIorcid{0009-0008-3419-0849},
	M.~S.~Anderson$^{39}$\BESIIIorcid{0009-0008-1550-2632}
% ...
	\\
	\vspace{0.2cm}
	(BESIII Collaboration)\\
	\vspace{0.2cm} {\it
	$^{1}$ Institute of High Energy Physics, Beijing 100049, People's Republic of China\\
% ...
$^{4}$ Budker Institute of Nuclear Physics SB RAS (BINP), Novosibirsk 630090, Russia\\
% ...
$^{31}$ INFN Sezione di Ferrara, (A)INFN Sezione di Ferrara, I-44122, Ferrara, Italy; (B)University of Ferrara, I-44122, Ferrara, Italy\\
% ...
$^{39}$ Johannes Gutenberg University of Mainz, Johann-Joachim-Becher-Weg 45, D-55099 Mainz, Germany\\
% ...
$^{65}$ State Key Laboratory of Particle Detection and Electronics, Beijing 100049, Hefei 230026, People's Republic of China\\
% ...
$^{78}$ University of Science and Technology of China, Hefei 230026, People's Republic of China\\
% ...
$^{81}$ University of Turin and INFN, (A)University of Turin, I-10125, Turin, Italy; (B)University of Eastern Piedmont, I-15121, Alessandria, Italy; (C)INFN, I-10125, Turin, Italy\\
	$^{82}$ Uppsala University, Box 516, SE-75120 Uppsala, Sweden\\
% ...
$^{88}$ Zhengzhou University, Zhengzhou 450001, People's Republic of China\\
	\vspace{0.2cm}
	$^{\dagger}$ Deceased\\
% ...
$^{c}$ Also at the Novosibirsk State University, Novosibirsk, 630090, Russia\\
% ...
	}
}

\begin{abstract}
Using $(10.087\pm0.044)\times10^9$ $J/\psi$ events and
$(2.712\pm0.014)\times10^9$ $\psi(3686)$ events collected by the
BESIII detector operating at the BEPCII collider, we report the first
observation of the hadronic decays of $J/\psi \to p \bar p K^0_S
K^0_S$ and $\psi(3686) \to p \bar p K^0_S K^0_S$, both with
statistical significance greater than $10\sigma$. Their branching
fractions are determined to be $\mathcal{B}(J/\psi \to p \bar p K^0_S
K^0_S)=(1.60 \pm 0.02 \pm 0.09)\times10^{-5}$ and
$\mathcal{B}(\psi(3686) \to p \bar p K^0_S K^0_S)=(3.93 \pm 0.24 \pm
0.34)\times10^{-6}$.  The ratio of their branching fractions is
$\mathcal{B}(\psi(3686) \to p \bar p K^0_S K^0_S)/\mathcal{B}(J/\psi
\to p \bar p K^0_S K^0_S)=(24.6 \pm 1.5 \pm 2.1)\%$, which deviates
from theoretical expectation by 4.6$\sigma$.  Here the first
uncertainties are statistical and the second systematic.  We have also
examined the $p\bar p$ invariant mass distributions in these decays,
and no significant enhancement around the $p \bar p$ near threshold is
found.  \end{abstract}

\maketitle

\section{Introduction}

Investigations of various decays of charmonia are important for the
understanding of Quantum Chromodynamics
(QCD)~\cite{ref::bes3-white-paper}. In perturbative QCD (pQCD), the
hadronic decays of both $J/\psi$ and $\psi(3686)$ are thought to
occur mainly via $c\bar{c}$ pair annihilation into three gluons,
with the branching fractions (BFs) proportional to the squares of the
charmonium wave functions at the origin~\cite{ref::aspect3}. This leads
to a ratio of BFs of $\psi(3686)$ and $J/\psi$ decays to light
hadrons:
\begin{equation}
\mathcal{Q}_{h}=\frac{\mathcal{B}_{\psi(3686)\to h}}{\mathcal{B}_{J/\psi\to h}} = \frac{\mathcal{B}_{\psi(3686)\to e^+e^-}}{\mathcal{B}_{J/\psi\to e^+e^-}} = 12.7\%,
\nonumber
\end{equation}
which is known as the pQCD 12\% rule.  A violation of this rule
was first reported by the Mark-II experiment in $\psi\to\rho\pi$ decay
in 1984~\cite{ref::aspect4}, and subsequently found in many other
decay channels. The discrepancy between the pQCD predictions and the
experimental measurements is known as the $\rho$-$\pi$ puzzle.
Different mechanisms have been proposed to explain this puzzle.
However, no model provides a self-consistent explanation for all
observed experimental results.  See Ref.~\cite{ref::aspect5} for a
review. %

In recent years, much progress has been made in the experimental
studies of multi-body decays of $J/\psi$ and $\psi(3686)$.  However,
knowledge of their decays into the $p\bar p K\bar K$ final states is
limited.  So far, only a few measurements of $J/\psi$ or $\psi(3686)$
decays into $p\bar p K^+K^-$ or $p\bar p \phi (\to K^+K^-)$, have been
reported
\cite{bes3:Jpsi_phippbar,DM2:Jpsi_phippbar,bes3:Psi2S_phippbar,CLEO:Psi2S_ppbarKK,bes3:Psi2S_phippbar},
yielding BFs around $10^{-6}$ to $10^{-5}$. Assuming isospin symmetry,
the BFs of $J/\psi \to p\bar p K^0_SK^0_S$ and $\psi(3686) \to p\bar p
K^0_SK^0_S$ would be one-fourth of those for $J/\psi \to p\bar p
K^+K^-$ and $\psi(3686) \to p\bar p K^+K^-$.  To date, no experimental
study of $\psi \to p\bar p K^0_SK^0_S$ (here $\psi$ stands for both
$J/\psi$ and $\psi(3686)$) has been reported.  The measurements of the
BFs of these two decays allow the test of the 12\% rule for $\psi\to
p\bar p K\bar K$ decays.

In addition, a structure was observed near the $p\bar p$ invariant-mass
threshold in the radiative decay of $J/\psi \to \gamma p\bar
p$~\cite{bes2:gammappbar,bes3:gammappbar}. Various theoretical models
have been proposed to explain this enhancement, but no definitive
conclusion has been reached~\cite{xwwang,dguo,Salnikov}. Studies of
the decays of $\psi\to p\bar p K^0_SK^0_S$ can provide additional
information of possible $p\bar p$ threshold structures.  Furthermore,
they also offer an opportunity to explore potential excited baryon
states in the $p K^0_S$ or $\bar p K^0_S$ systems.

In this paper, we present the first observations and BF measurements
of the hadronic decays of $J/\psi \to p \bar p K^0_S K^0_S$ and
$\psi(3686) \to p \bar p K^0_S K^0_S$, by analyzing
$(10.087\pm0.044)\times10^9$ $J/\psi$ events~\cite{ref::jpsi-num-inc}
and $(2.712\pm0.014)\times10^9$ $\psi(3686)$
events~\cite{ref::psip-num-inc} collected by the BESIII detector.

\section{BESIII DETECTOR AND MONTE CARLO SIMULATION}
\label{sec:BES}

The BESIII detector~\cite{ref::detector} records symmetric $e^+e^-$ collisions provided by the BEPCII storage ring~\cite{ref::collider}
in the center-of-mass energy range from 1.84 to 4.95~GeV,
with a peak luminosity of $1.1 \times 10^{33}\;\text{cm}^{-2}\text{s}^{-1}$
achieved at $\sqrt{s} = 3.773\;\text{GeV}$.

The cylindrical core of the BESIII detector covers 93\% of the full solid angle and consists of a helium-based
 multilayer drift chamber~(MDC), a plastic scintillator time-of-flight
system~(TOF), and a CsI(Tl) electromagnetic calorimeter~(EMC),
which are all enclosed in a superconducting solenoidal magnet
providing a 1.0~T magnetic field.
The solenoid is supported by an
octagonal flux-return yoke with resistive plate counter muon
identification modules interleaved with steel.
The charged-particle momentum resolution at $1~{\rm GeV}/c$ is
$0.5\%$, and the
${\rm d}E/{\rm d}x$
resolution is $6\%$ for electrons
from Bhabha scattering. The EMC measures photon energies with a
resolution of $2.5\%$ ($5\%$) at $1$~GeV in the barrel (end cap)
region. The time resolution in the TOF barrel region is 68~ps, while
that in the end cap region was 110~ps.
The end-cap TOF system was upgraded in 2015 using multi-gap resistive plate chamber technology, providing a time resolution of 60~ps~\cite{Tof1,Tof2,Tof3}.
This benefit 86\% of the $J/\psi$ sample and 85\% of the $\psi(3686)$ sample used in the analysis.

Simulated data samples produced with {\sc geant4}-based~\cite{Geant4}
Monte Carlo (MC) software, which includes the geometric description of
the BESIII detector and the detector response, are used to determine
detection efficiencies and to estimate background contributions. The simulation
models the beam energy spread and initial state radiation (ISR) in the
$e^+e^-$ annihilations with the generator {\sc kkmc}~\cite{Jadach01}.
Inclusive MC samples containing about $10^{10}~J/\psi$ events and
about $2.7\times10^9~\psi(3686)$ events include the production of the
$J/\psi$ or $\psi(3686)$ resonance, the $J/\psi$ ISR production, and the continuum processes incorporated in {\sc kkmc}~\cite{Jadach01}.  All particle decays are modeled with {\sc evtgen}~\cite{Lange01} using BFs either taken from the Particle Data Group (PDG)~\cite{ref::pdg2022}, when available, or otherwise
estimated with {\sc lundcharm}~\cite{Lundcharm00}.  Final state
radiation from charged final state particles is incorporated using
{\sc photos}~\cite{PHOTOS}.  Exclusive MC samples of $5\times10^5$
$J/\psi$ and $\psi(3686)\to p \bar p K_S^0 K_S^0$ events with subsequent
decay $K_S^0\to \pi^+\pi^-$, based on the phase space (PHSP) model,
are generated to determine the detection efficiencies.

\section{EVENT SELECTION}
\label{sec:selection}

Candidates for the decays $J/\psi \to p \bar p K^0_S K^0_S$ and
$\psi(3686) \to p \bar p K^0_S K^0_S$ are required to have at least six charged
tracks.  These tracks detected in the MDC are required to be
within a polar angle ($\theta$) range of $|\cos\theta|<0.93$, where
$\theta$ is defined with respect to the $z$-axis, which is the
symmetry axis of the MDC. The distance of closest approach to the
interaction point (IP), except for $\pi^\pm$s from $K^0_S$ decays,
must be less than 10\,cm along the $z$-axis, $|V_{z}|$, and less than
1\,cm in the transverse plane, $|V_{xy}|$.  Particle
identification~(PID) for charged tracks combines measurements of the
(d$E$/d$x$) and the flight time in the TOF to compute likelihoods
$\mathcal{L}(h)~(h=p,K,\pi)$ for each hadron $h$ hypothesis. Charged
tracks are identified as protons when the proton hypothesis has the
largest likelihood.

The $K^0_S$ candidates are reconstructed from two oppositely charged
tracks that satisfy $|V_{z}|<20$~cm. No constraint on $|V_{xy}|$ is
required, and the tracks are assigned as $\pi^\pm$ without imposing
further PID criteria.  Secondary vertex fits are performed, and the
decay lengths of the $K^0_S$ candidates must be more than twice the
resolution of the vertex positions.

A four-momentum conservation constraint (4C) kinematic fit is applied
under the hypothesis of $e^+e^-\to p\bar p K^0_SK^0_S$, where two
$K^0_S$ candidates are included as virtual particles with information
derived from the secondary vertex fits. In each event, if more than
one combination survives, the one with the smallest $\chi_{\rm
4C}^{2}$ from the 4C kinematic fit is retained.  The $\chi_{\rm
4C}^{2}$ is required to be less than 100,
which is based on the optimization with the Figure of Merit (FOM)
defined as \begin{equation} \mathrm{FOM} =
\frac{\mathit{S}}{\sqrt{\mathit{S}+\mathit{B}}}.  \end{equation} Here
$S$ denotes the number of events from the signal MC samples, normalized
according to our measured BFs, and $B$ denotes the number of
background events from the inclusive MC samples, normalized to the size of the data
sample.

The potential background components from $J/\psi$ and $\psi(3686)$
decays are studied by analyzing individual inclusive MC samples with TopoAna tool~\cite{ref::topoana}. No
significant peaking backgrounds are observed.  To suppress the
background contribution from the decay $\psi\to p K^{*-}\bar\Lambda + c.c.$, the
invariant mass of all $p \pi^-$ and $\bar p \pi^+$ combinations are
required to satisfy $|M_{p \pi}-M_\Lambda| > 0.015$ GeV/$c^2$, where
$M_\Lambda$ is the nominal $\Lambda$ mass~\cite{ref::pdg2022}. Furthermore, to estimate the
continuum (non-resonant) background, we analyse the data at $\sqrt s =$
3.08~GeV and 3.65~GeV collected by the BESIII detector with the integrated
luminosities of 167.06~pb$^{-1}$~\cite{ref::jpsi-num-inc} and
445.5~pb$^{-1}$~\cite{ref::psip-num-inc}, respectively.

\begin{figure*}[htbp]
  \begin{center}
  \includegraphics[width=0.95\textwidth] {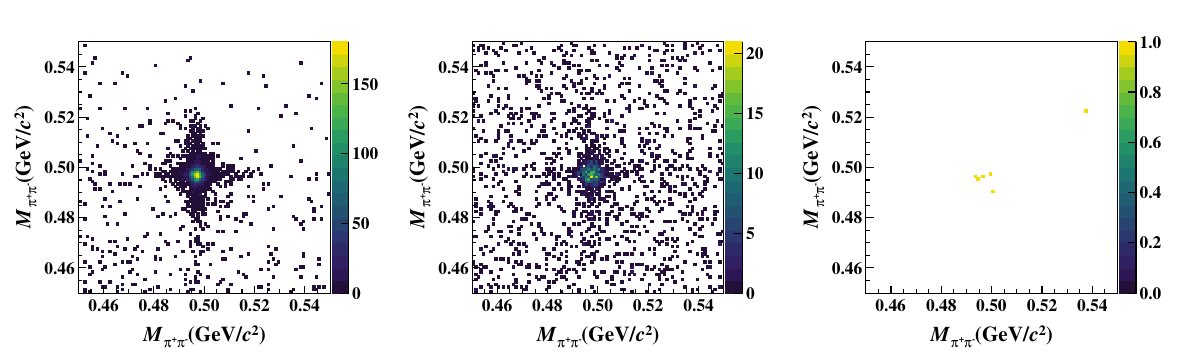}
  \caption{The $M_{\pi^+\pi^-(1)}$ versus $M_{\pi^+\pi^-(2)}$ distributions of the accepted candidates for
  $e^+e^- \to p \bar p K^0_S K^0_S$ from the (left) $J/\psi$, (middle) $\psi(3686)$ and (right) $\sqrt{s}=3.65$~GeV continuum data samples.}
  \label{fig:Scatter}
  \end{center}
\end{figure*}

Figure~\ref{fig:Scatter} shows the invariant masses of the two $\pi^+\pi^-$
combinations, $M_{\pi^+\pi^-(1)}$ versus
$M_{\pi^+\pi^-(2)}$, of the accepted candidates for
  $e^+e^- \to p \bar p K^0_S K^0_S$  from the $J/\psi$, $\psi(3686)$ and continuum data samples.
  Clusters around both $K^0_S$ masses are observed.

\begin{figure*}[htbp]
  \begin{center}
\includegraphics[width=0.32\textwidth] {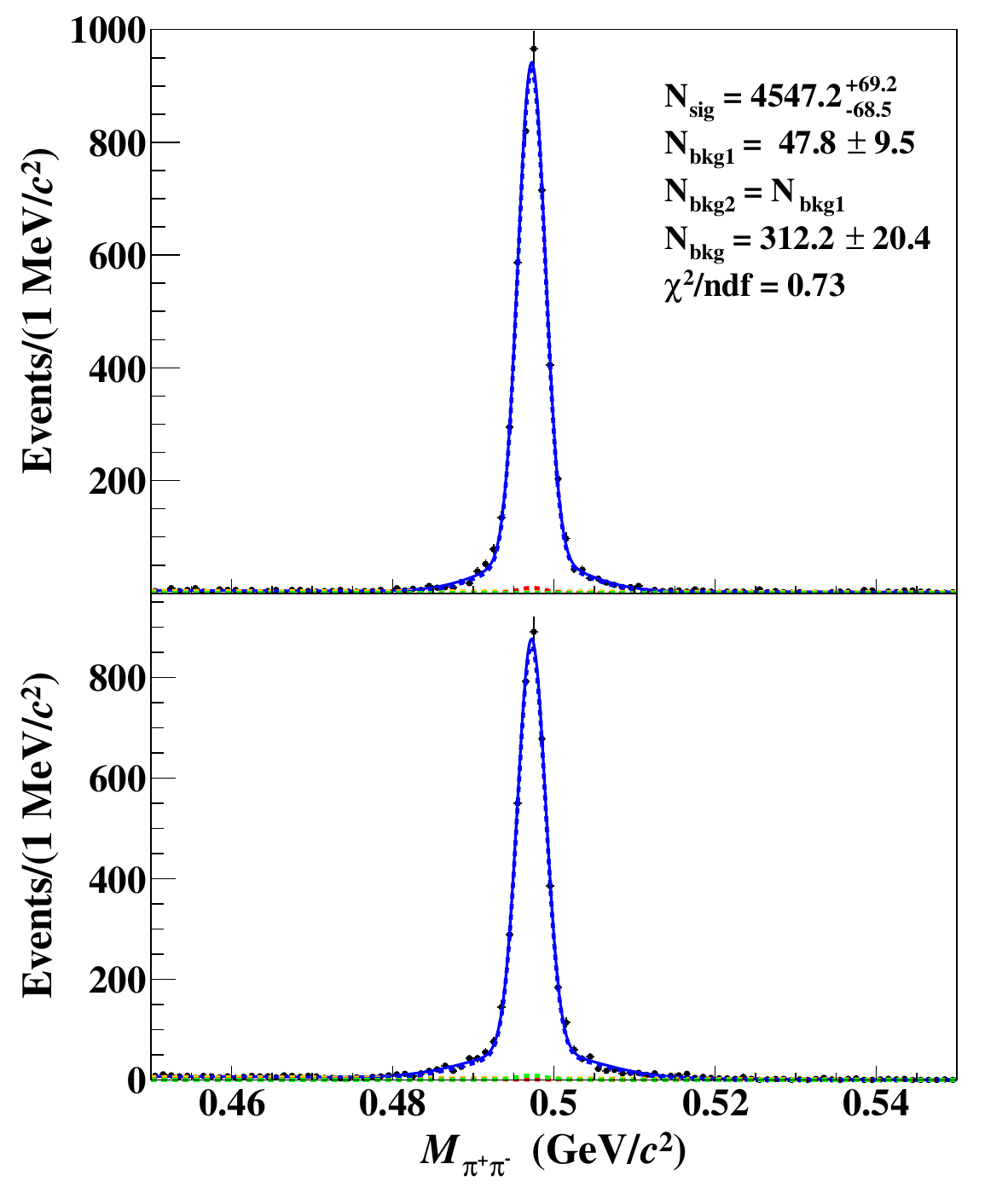}
\includegraphics[width=0.32\textwidth] {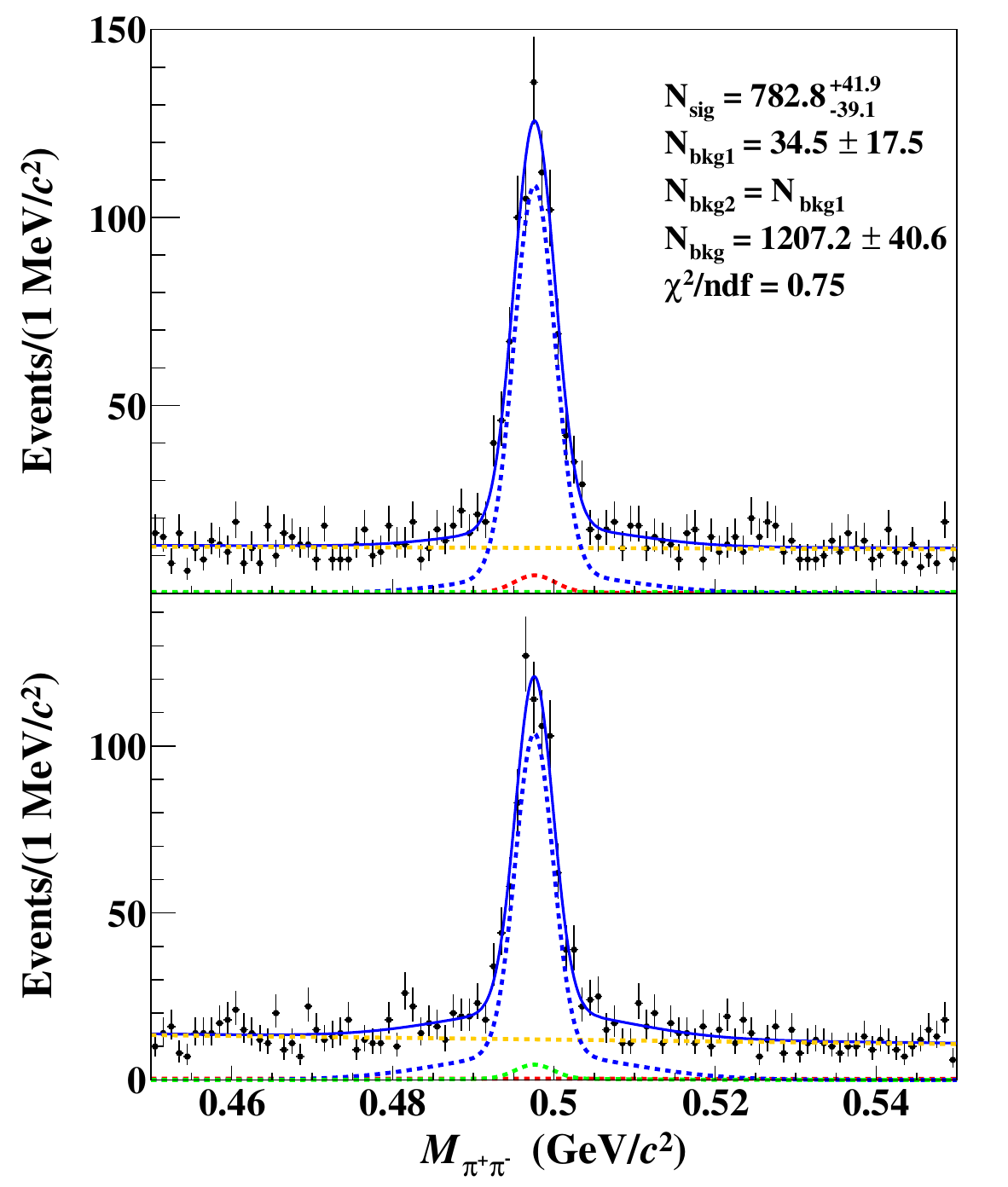}
\includegraphics[width=0.32\textwidth] {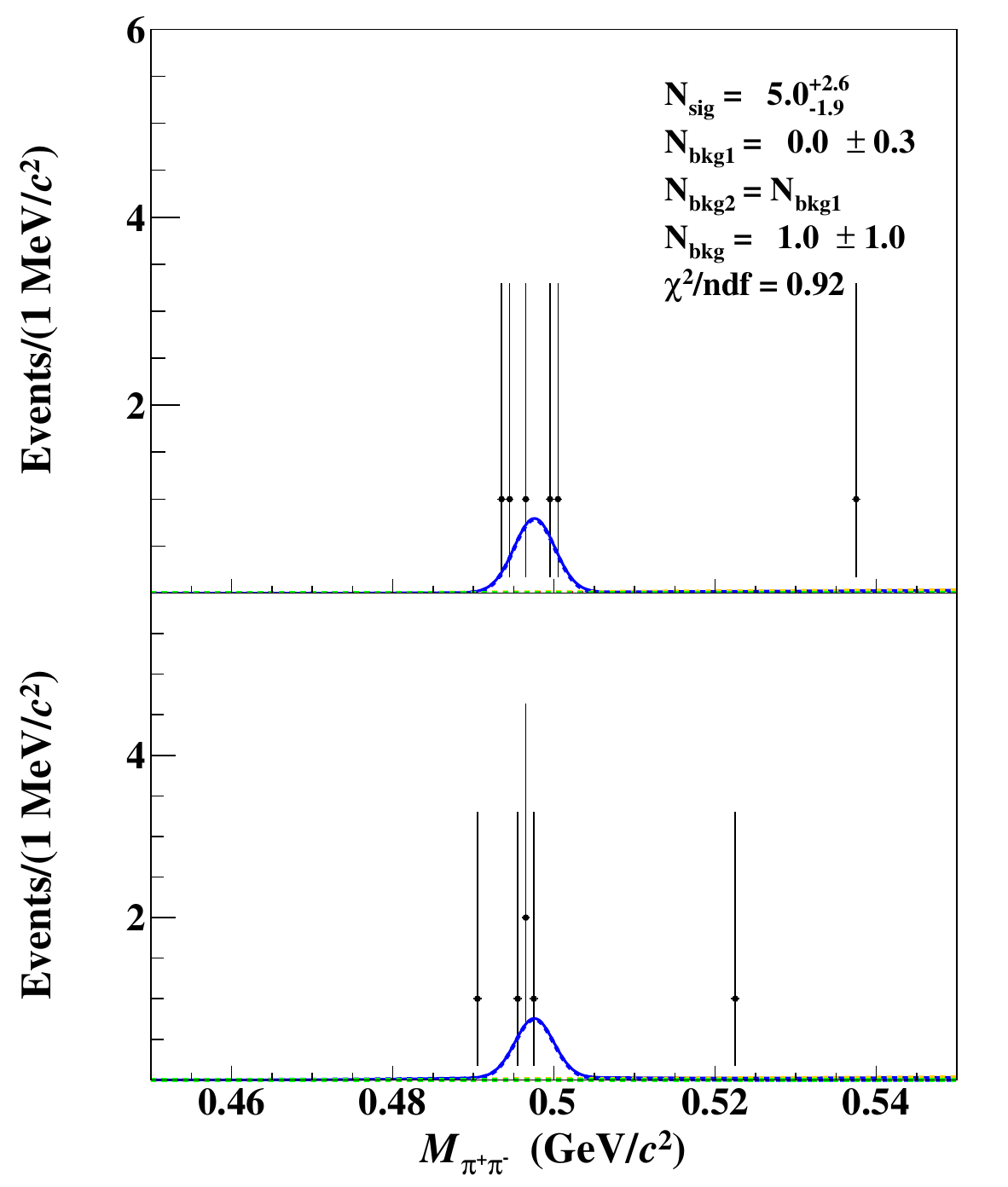}
  \caption{The projections of the 2D fits on (top) $M_{\pi^+\pi^-(1)}$ and (bottom) $M_{\pi^+\pi^-(2)}$ of the accepted candidates for the process $e^+e^- \to p \bar p K^0_S K^0_S$ from the (left) $J/\psi$, (middle) $\psi(3686)$ and (right) continuum data samples.
  The black dots with error bars are the data, the blue solid curves are the total fit, the blue dash lines are the signal candidates. The red and green dashed curves are the background candidates with single $K_S^0$ (bkg1 and bkg2), respectively, and the yellow dashed lines are the continuum background (bkg).
  }
  \label{fig:2Dfit}
  \end{center}
\end{figure*}

To determine the signal yields, unbinned maximum likelihood
fits are performed on the two-dimensional (2D) distributions of
$M_{\pi^+\pi^-(1)}$ versus $M_{\pi^+\pi^-(2)}$ of the accepted
candidates in data. In the fits, the $K^0_S$ signal shape is described
by a double Gaussian with free parameters and the combinatorial
background shape is described by a first-order polynomial function. The
2D probability density functions of signal, background~1 (bkg1), background~2 (bkg2) and continuum background~(bkg) are
construncted as follows:
\begin{itemize}
	\item signal: $\mathcal{S}(M_{\pi^+\pi^-(1)})\times \mathcal{S}(M_{\pi^+\pi^-(2)})$,
	\item bkg1: $\mathcal{S}(M_{\pi^+\pi^-(1)})\times \mathcal{B}(M_{\pi^+\pi^-(2)})$, 
	\item bkg2: $\mathcal{B}(M_{\pi^+\pi^-(1)})\times \mathcal{S}(M_{\pi^+\pi^-(2)})$, 
	\item bkg: $\mathcal{B}(M_{\pi^+\pi^-(1)})\times
	\mathcal{B}(M_{\pi^+\pi^-(2)})$, \end{itemize} where
	$\mathcal{S}$ and $ \mathcal{B}$ denote the $K_S^0$ signal and
	background fitted functions, respectively.  The fit results
	are shown in Figure~\ref{fig:2Dfit}. From these fits, the
	signal yields of $e^+e^-\to p \bar p K^0_S K^0_S$ from the
	$J/\psi$, $\psi(3686)$ and $\sqrt{s}=3.65$~GeV continuum data
	samples are listed in Table~\ref{tab:BFs}, where the
	uncertainties are statistical only. The statistical
	significances for the former two cases are greater than
	$10\sigma$ for each decay estimated by comparing
	the fit likelihood values with and without each signal
	component separately.  No event passes selection in the data sample taken at $\sqrt{s}=3.08$~GeV. Thus, the contribution of continuum (non-resonant) events on $J/\psi$ decay is negligible.

The net signal yields of $\psi \to p \bar p K^0_S K^0_S$ are
determined via \begin{equation} N^{\rm net}_{\psi}=N^{\rm
obs}_{\psi}-f_{\rm co} N^{\rm obs}_{\rm con}.  \end{equation} Here,
$f_{\rm co} = \frac{\mathcal L_{3.686}} {\mathcal
L_{3.65}}\times\frac{3.65^{2n}}{3.686^{2n}}\times\frac{\epsilon^{\rm
QED}_{3.686}}{\epsilon^{\rm QED}_{3.650}}$ for $\psi(3686)$ and 
$f_{\rm co} = \frac{\mathcal L_{3.097}} {\mathcal
L_{3.08}}\times\frac{3.08^{2n}}{3.097^{2n}}\times\frac{\epsilon^{\rm
QED}_{3.097}}{\epsilon^{\rm QED}_{3.08}}$ for $J/\psi$
is the normalization factor
taking into account the difference of luminosity, center-of-mass
energy, and detection efficiency for the continuum production process
at different energy points~\cite{ref::interference}.  We take $n=1$
and calculate $f_{\rm co}$ to be 9.4 for the $\psi(3686)$ decay and
2.9 for the $J/\psi$ decay.  The resulting net signal yields of $\psi
\to $ $p \bar p K^0_S K^0_S$ are listed in Table~\ref{tab:BFs}.

\section{Branching fraction}

\begin{figure*}[htbp]
  \begin{center}
  \includegraphics[width=0.96\textwidth] {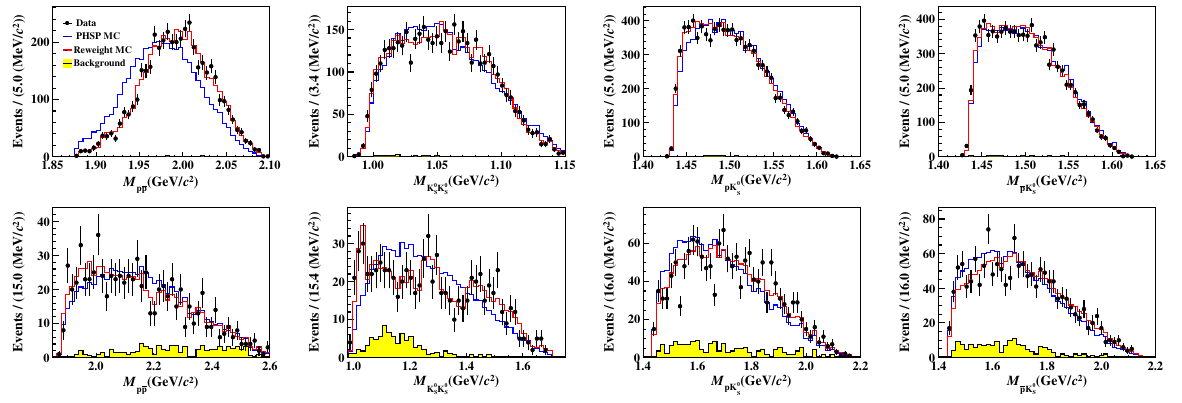}
\caption{The distributions of two-body invariant masses of the
accepted candidates for $J/\psi \to p \bar p K^0_S K^0_S$ (top row)
and $\psi(3686) \to p \bar p K^0_S K^0_S$ (bottom row).  Events are
required to be within $|M_{\pi^+\pi^-}-0.498|<0.012$~GeV/$c^2$, and
each event appears in the histograms twice for $M_{p K^0_S}$, and
$M_{\bar p K^0_S}$. The black dots with error bars are the data, the blue and red solid curves are the phase space and re-weighted MC events, the yellow histograms represent the background candidates.}
  \label{fig:compare}
  \end{center}
\end{figure*}

To determine the detection efficiencies, the signal MC samples of
$J/\psi$ and $\psi(3686) \to p \bar p K^0_S K^0_S$ are re-weighted
based on the 2D distribution of $M_{p\bar p}$ versus $M_{K^0_SK^0_S}$
using data sample. The comparisons of the distributions of two-body invariant
masses of the accepted candidates for $J/\psi \to p \bar p K^0_S
K^0_S$ and $\psi(3686) \to p \bar p K^0_S K^0_S$ are shown in
Figure~\ref{fig:compare}.
The $M_{p \bar p}$ spectra in the first column of
Figure~\ref{fig:compare} shows no clear enhancement structure near the
$p \bar p $ threshold.  The detection efficiencies for $J/\psi \to p
\bar p K^0_S K^0_S$ and $\psi(3686) \to p \bar p K^0_S K^0_S$ are
listed in Table~\ref{tab:BFs}.

\label{sec:mc}
The BFs of the processes $J/\psi \to p \bar p K^0_S K^0_S$ and $\psi(3686) \to p
\bar p K^0_S K^0_S$ are calculated as \begin{equation} {\mathcal
B}_{\psi\to p \bar p K^0_S K^0_S}=\frac{N^{\rm
net}}{N_{\psi}\cdot\epsilon \cdot {\mathcal B}^2_{K^0_S\to
\pi^+\pi^-}}, \end{equation} where $N_{\psi}$ is the number of
$J/\psi$ or $\psi(3686)$ events in data, $\epsilon$ is the detection
efficiency, and $\mathcal B_{K^0_S\to \pi^+\pi^-}$ is the BF of
$K^0_S\to \pi^+\pi^-$ taken from the PDG~\cite{ref::pdg2022}.
Substituting $N^{\rm net}$, $N_{\psi}$, $\epsilon$, and $\mathcal
B_{K^0_S\to \pi^+\pi^-}$ into this equation, we obtain the BFs of
$J/\psi \to p \bar p K^0_S K^0_S$ and $\psi(3686) \to p \bar p K^0_S
K^0_S$, which are listed in Table~\ref{tab:BFs}.

\begin{table}[htbp]\small
	\caption{The net signal yields, the detection efficiencies, and
 the obtained BFs of $J/\psi \to p \bar p K^0_S K^0_S$ and $\psi(3686) \to p \bar p K^0_S K^0_S$, where the  uncertainties are statistical only.}
	\centering
		\begin{tabular}{l|c|c}
   \hline
   \hline
   & $J/\psi\to p \bar p K_S^0 K_S^0$ & $\psi(3686)\to p \bar p K_S^0 K_S^0$ \\ \hline
   $N^{\rm obs}$         & $4547.4\pm68.8$& $782.8\pm40.5$\\
   $N_{\rm con}^{\rm obs}$         & 0           & $5.0\pm2.2$   \\
   $N^{\rm net}$         & $4547.4\pm68.8$    & $735.8\pm 45.5$   \\
   $N_{\psi}$       & $(1.0087\pm0.0044)\times10^{10}$ & $(2.712\pm0.014)\times10^9$        \\
   $\epsilon$            & $(5.88 \pm 0.03)\%$& $  (14.41 \pm 0.07)\%$\\
  $\mathcal{B}$   & $(1.60 \pm 0.02)\times10^{-5}$& $(3.93\pm0.24)\times10^{-6}$\\
   \hline
   \hline
         \end{tabular}
	\label{tab:BFs}
\end{table}

\section{SYSTEMATIC UNCERTAINTY}
\label{sec:systematics}

The systematic uncertainties in the BF measurements, that originate from several sources, are listed in Table~\ref{tab:Sys} and discussed below.

\begin{table}[htbp]
	\centering
  	\caption{Relative systematic uncertainties (in \%) in the BF measurements. The uncertainty related to the continuum is negligible for $J/\psi$.}
  \begin{tabular}{lcc}
  	\hline
  	\hline
    Source                         & $J/\psi$ & $\psi(3686)$ \\
    \hline
  	$p(\bar p)$ tracking                 & 2.0& 2.0\\
    $p(\bar p)$ PID                      & 2.0& 2.0\\
    $K_{S}^{0}$ reconstruction           & 3.0& 3.0\\
    $N_{\psi}$                           & 0.4& 0.5\\
    $M_{\pi^+\pi^-}$ fit                 & 2.5 & 3.3 \\
    MC model                             &  1.7  & 0.2\\
    $\Lambda$ veto                       &  2.2  &  2.1 \\
    $f_{\rm co}$ factor                       & ...&  1.0 \\
    MC statistics                        & 0.5& 0.5 \\
    $\mathcal{B}(K_{S}^{0}\to\pi^+\pi^-)$ & 0.2& 0.2\\
    QED interference                     & 0.3 & 6.3\\   \hline
    Total                               &  5.6  &  8.6 \\
  	\hline
  	\hline
  \end{tabular}
  \label{tab:Sys}
\end{table}

The systematic uncertainties of $p(\bar{p})$ tracking or PID are
estimated with the control sample of $J/\psi \to p\bar p
\pi^+\pi^-$ to be 1.0\% for each case~\cite{ref::tracking}.

The systematic uncertainty of $K^0_S$ reconstruction, including
tracking efficiency, $K^0_S$ mass window, vertex fit and second vertex
fit, are estimated by using control samples of $J/\psi \to
K^0_SK^-\pi^++c.c.$ and $J/\psi \to \phi K^0_SK^-\pi^++c.c.$.  The
systematic uncertainty of the $K^0_S$ reconstruction is assigned to be
1.5\%~\cite{ref::KS-recon}.

The numbers of $J/\psi$ and $\psi(3686)$ events in data are
measured to be $(10.087\pm0.044)\times10^9$ $J/\psi$ and
$(2.712\pm0.014)\times10^9$, respectively, by using inclusive hadronic
events~\cite{ref::jpsi-num-inc,ref::psip-num-inc}.  Their corresponding uncertainties are 0.4\% and 0.5\%.

The systematic uncertainties arising due to the 2D fit to the $M_{\pi^+\pi^-(1)}$
versus $M_{\pi^+\pi^-(2)}$ distribution are considered in two parts.
The uncertainties from the background shape are estimated by replacing
the order of the nominal polynomial function from first order to second
order. The systematic uncertainties from the signal shapes are estimated
by using Breit-Wigner function convolved with a Gaussian resolution function. The observed differences in the measured branching fractions are assigned as the corresponding systematic uncertainties. Adding these two effects in quadrature yields the systematic uncertainties of 2.5\% and 3.3\% for the $J/\psi$ and
$\psi(3686)$ decays, respectively.

To estimate the systematic uncertainties from the MC model, we change the
bin division in the re-weighting procedure from 16 bins to 12 and 20 bins.
The maximum changes of the signal efficiencies, 1.7\% and 0.2\%, are
assigned as the systematic uncertainties for the
$J/\psi$ and $\psi(3686)$ decays, respectively.

The systematic uncertainty due to $\Lambda$ veto is estimated by
changing the $\Lambda$ range requirement from 9~MeV/$c^2$ to 21~MeV/$c^2$. The maximum changes of the measured BFs, 2.2\% for $J/\psi$
and 2.1\% for $\psi(3686)$, are taken as the corresponding systematic
uncertainties.

The systematic uncertainty of the normalization factor $f_{\rm co}$ for
$\psi(3686)$ is considered in two parts. The statistical error of the
MC-determined ratio $\epsilon^{\rm QED}_{3.686} / \epsilon^{\rm
QED}_{3.650}$ is 1.0\%. The uncertainty due to the unknown energy
dependence of cross section is estimated by switching the power $n$ of
$\sqrt{s}$ in the $f_{\rm co}$ calculation from 1 to 3. The change on the BF
is taken as the systematic uncertainty. Adding the two parts in
quadrature gives the uncertainty for $\psi(3686)$.

The systematic uncertainties due to limited statistics of the signal MC samples
are both 0.5\% for the $J/\psi \to p \bar p K^0_S K^0_S$ and $\psi(3686) \to p \bar p K^0_S K^0_S$ processes.

The uncertainty of the $K^0_S\to \pi^+\pi^-$ BF is 0.1\% taken from the PDG~\cite{ref::pdg2022}.

The BF is determined without considering the interference between the $\psi(3686)$ and continuum amplitudes.  The uncertainty due to this effect is estimated by introducing an interference term~\cite{ref::interference}. The relative change of the signal yield due to the unknown interference angle, which is 6.3\% for
$\psi(3686)$, is taken as the systematic uncertainty.  For $J/\psi$,
the QED contribution is negligible, and the corresponding systematic
uncertainty is estimated by using 3.65~GeV data to calculate the QED
contribution. The effect on BF is 0.3\%, which is taken as the systematic
uncertainty.

For each signal decay, the total systematic uncertainty is obtained by
adding all systematic uncertainties in quadrature. While calculating
the ratio $\mathcal{B}_{\psi(3686) \to p \bar p K^0_S
K^0_S}/\mathcal{B}_{J/\psi \to p \bar p K^0_S K^0_S}$, the
uncertainties due to tracking, PID, $K_S^0$ reconstruction and cited
$\mathcal{B}(K_S^0 \to \pi^+\pi^-)$ cancel.

\section{Summary}

Based on the analyses of $(10.087\pm0.044)\times10^9$ $J/\psi$ events and $(2.712\pm0.014)\times10^9$ $\psi(3686)$ events collected by the BESIII detector, the hadronic decays of $J/\psi \to p \bar p K^0_S K^0_S$ and $\psi(3686) \to p \bar p K^0_S K^0_S$ are observed for the first time with statistical significances greater than $10\sigma$. We find
\begin{equation}
\mathcal{B}_{J/\psi \to p \bar p K^0_S K^0_S}= (1.60 \pm 0.02 \pm 0.09)\times10^{-5}
\nonumber
\end{equation}
and
\begin{equation}
\mathcal{B}_{\psi(3686) \to p \bar p K^0_S K^0_S}= (3.93 \pm 0.24 \pm 0.34)\times10^{-6},
\nonumber
\end{equation}
where the first uncertainties are statistical and the second are systematic.
Furthermore, we obtain the ratio
\begin{equation}
\mathcal{Q}_{p \bar p K^0_S K^0_S}=\frac{\mathcal{B}_{\psi(3686) \to p \bar p K^0_S K^0_S}}{\mathcal{B}_{J/\psi \to p \bar p K^0_S K^0_S}} = (24.6 \pm 1.5 \pm 2.1)\%,
\nonumber
\end{equation}
which differs from the theoretical expectation of the 12\% rule by 4.6$\sigma$. Combining with $\mathcal{B}_{\psi(3686)\to p\bar p K^+K^-}$ $=(2.7\pm0.7) \times10^{-5}$ from Ref.~\cite{CLEO:Psi2S_ppbarKK}, we determine
\begin{equation}
\frac{\mathcal{B}_{\psi(3686)\to p\bar p K^+K^-}}{4\cdot \mathcal{B}_{\psi(3686) \to p \bar p K^0_S K^0_S}} = 1.7 \pm 0.5,
\nonumber
\end{equation}
which is consistent with the prediction based on isospin symmetry within $1.4\sigma$.
In addition, we have examined the $p\bar p$ invariant mass spectra in these decays, and no clear enhancement structure around the $p\bar p$ near threshold is found.

\textbf{Acknowledgement}

The BESIII Collaboration thanks the staff of BEPCII (https://cstr.cn/31109.02.BEPC) and the IHEP computing center for their strong support. This work is supported in part by National Key R\&D Program of China under Contracts Nos. 2025YFA1613900, 2023YFA1606000, 2023YFA1606704; National Natural Science Foundation of China (NSFC) under Contracts Nos. 11635010, 11935015, 11935016, 11935018, 12025502, 12035009, 12035013, 12061131003, 12192260, 12192261, 12192262, 12192263, 12192264, 12192265, 12221005, 12225509, 12235017, 12342502, 12361141819; the Chinese Academy of Sciences (CAS) Large-Scale Scientific Facility Program; the Strategic Priority Research Program of Chinese Academy of Sciences under Contract No. XDA0480600; CAS under Contract No. YSBR-101; 100 Talents Program of CAS; The Institute of Nuclear and Particle Physics (INPAC) and Shanghai Key Laboratory for Particle Physics and Cosmology; ERC under Contract No. 758462; German Research Foundation DFG under Contract No. FOR5327; Istituto Nazionale di Fisica Nucleare, Italy; Knut and Alice Wallenberg Foundation under Contracts Nos. 2021.0174, 2021.0299, 2023.0315; Ministry of Development of Turkey under Contract No. DPT2006K-120470; National Research Foundation of Korea under Contract No. NRF-2022R1A2C1092335; National Science and Technology fund of Mongolia; Polish National Science Centre under Contract No. 2024/53/B/ST2/00975; STFC (United Kingdom); Swedish Research Council under Contract No. 2019.04595; U. S. Department of Energy under Contract No. DE-FG02-05ER41374

\end{document}